# Interplay of Electric Dipole Spin Resonance and Multilevel Landau-Zener Interference in p-Type Silicon Quantum Dots

Sayyid Irsyadul Ibad[1,*], Yusaku Suzuki[1], Masahiro Tadokoro[1], Tokio Futaya[1], Shimpei Nishiyama[1,2], Kimihiko Kato[2], Shigenori Murakami[2], Takahiro Mori[2], Raisei Mizokuchi[1], Jun Yoneda[3] and Tetsuo Kodera[1,*]

[1]Department of Electrical and Electronic Engineering, Institute of Science Tokyo, Tokyo 152-8550, Japan
[2]National Institute of Advanced Industrial Science and Technology (AIST), Ibaraki 305-8569, Japan
[3]Department of Advanced Materials Science, University of Tokyo, Chiba 277-8561, Japan

**ABSTRACT**. In this work, we examine microwave responses of the Pauli spin blockade (PSB) leakage current through a p-type silicon double quantum dot. We observe more than the expected two resonance lines with the main resonance line exhibits both positive and negative peaks as a function of the magnetic field, corresponding to enhancement and suppression of the PSB leakage current, respectively. We attribute the observed spectra to the interplay between two spin rotation mechanisms: spin-orbit-mediated electric dipole spin resonance (EDSR) and multilevel Landau-Zener (MLLZ) interference, both of which are present in electrically driven devices with strong spin-orbit coupling (and enhanced in the vicinity of orbital level crossings). A numerical simulation taking into account both mechanisms show agreement with the experimental results. While these unconventional spectral behaviours can be readily suppressed away from the orbital level crossing or in devices with weak spin-orbit coupling, our study showcases the potential complexity of spin-rotating mechanisms for electrically driven spin qubits.

## I. INTRODUCTION.

Semiconductor spins are promising candidates for realizing quantum computers due to their compatibility with industrial fabrication, which enables dense integration of qubits with classical control electronics at cryogenic temperatures [1–4]. Recently, hole spin qubits have gained attention, due to their inherent strong spin-orbit coupling (SOC). This enables fast and fully electrical control of spin qubits via electric dipole spin resonance (EDSR) without the need for additional structures such as micromagnets, microwave striplines, or other auxiliary components, thereby reducing the complexity of scaling up qubit devices. It has been shown that hole spin qubits can be controlled via EDSR with Rabi frequencies reaching hundreds of megahertz [5–8], surpassing typical values of only a few megahertz reported for the electron-based qubits [9–12].

Pauli spin blockade (PSB) in multiple quantum dots is commonly utilized as a spin-to-charge conversion technique for spin state initialization and readout. This approach is advantageous because it relies on blocked versus unblocked spin configurations and does not require large Zeeman splitting, making it more robust than energy-resolved tunneling schemes, enabling high-fidelity initialization and readout even at elevated temperatures [10,13–18]. Moreover, PSB can be utilized as a tool for investigating spin dynamics [18–21]. For example, peaks arising from the microwave responses of the PSB leakage transport current provide information on the energy level structures of the system [21–23]. These peaks can exhibit various line shapes that could offer insights on the underlying physics of electrically driven quantum dots [24–26].

Recent theoretical studies suggest that the various line shapes can be attributed to either SOC mediated EDSR or multi-level Landau-Zener (MLLZ) interference [27], both of which occur when quantum dots are electrically driven in the presence of strong SOC. However, these mechanisms have so far been treated separately, motivating investigation of their possible coexistence. In the EDSR process, an AC electric field modulates the wavefunction, and this motion is then translated into an effective oscillating magnetic field by SOC [28–31]. If the oscillation is on resonance with the Zeeman energy, it can induce spin rotation between two spin states. On the other hand, in the MLLZ interference process, the alternating electric field oscillates the quantum dots energy detuning, and a resonant oscillation can induce spin transitions between two spin states mediated by a third energy level [32,33]. Thus, if the two mechanisms coexist, the interplay between them may lead to complex spin dynamics, which is crucial to study for achieving precise spin qubit control.

*Contact authors:
sayyid.i.5937@m.isct.ac.jp
kodera.t.a173@m.isct.ac.jp

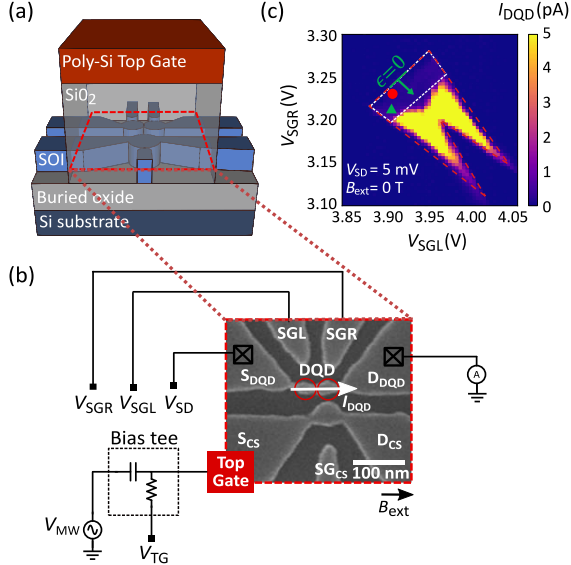

FIG 1. (a) Schematic of the silicon double quantum dot (DQD) device used in this work. The quantum dots are physically defined by patterning the silicon-on-insulator (SOI) layer above the buried oxide. A polysilicon top gate is deposited on the silicon oxide to accumulate a two-dimensional hole gas (2DHG) at the SOI/SiO$_2$ interface. (b) Schematic of the experimental setup and a scanning electron microscopy image of the quantum dot device. The device consists of a DQD and a proximate charge sensor (CS), which is not used in this experiment. An external magnetic field is applied parallel to the DQD channel. (c) Bias triangles (outlined by red dashed lines) and the Pauli spin blockade (PSB) region (outlined by white dashed lines), measured with a source–drain bias voltage of 5 mV at 0 T. The red circle and green triangle markers indicate the low- and high-detuning points used in the spin resonance experiment, respectively.

In this work, we show the evidence of the coexistence of both mechanisms from microwave responses of the PSB leakage current in a p-type silicon double quantum dot (DQD). We observe more than the expected two resonance lines with one of them exhibiting an asymmetric current peak-and-dip that corresponds to the lifting and enhancement of the PSB, respectively. These phenomena show a strong dependence on the DQD level detuning. We attribute these phenomena to the interplay between EDSR and MLLZ interference, where the peak is due to EDSR process while the multiple dips due to MLLZ interference. In addition, the existence of asymmetric peak-and-dip can be understood as due to the competition between both mechanisms.

We perform numerical simulations considering both EDSR and MLLZ interference processes and qualitatively reproduce the experimental results. We furthermore confirm that the experimental results cannot be fitted well when one of the processes is neglected in the simulation model. We thus conclude that EDSR and MLLZ interference coexist and the interplay between them plays an essential role in the observation of multiple spin resonance lines with asymmetric peak-and-dip shape. Such interplay can complicate the dynamics of electrically driven spin qubits, and understanding these mechanisms may prove useful for optimizing spin qubit control.

## II. EXPERIMENTAL RESULTS

The experiment was performed using a p-type, physically defined silicon DQD at 300 mK, as shown in Fig. 1(a). The DQD is physically formed by patterning an undoped 40-nm-thick silicon-on-insulator (SOI) layer using electron beam lithography [34–40]. It is equipped with a polysilicon top gate with an area of 0.09 μm², positioned above the DQD and separated from the SOI layer by SiO$_2$ with thickness of ~65 nm. A negative DC voltage and an electric microwave signal is applied to the top gate to accumulate holes in the DQD and to perform spin manipulation, respectively, see Fig. 1(b). Positive bias voltages are applied to the two side gates, SGL and SGR to tune the DQD potential. We characterize the device by measuring the transport current under a source-drain bias voltage of 5 mV.

To achieve qubit control, we tune the gate voltages of the DQD device to obtain the bias triangles shown in Fig. 1(c). The current is suppressed on the base of the triangles, which indicates a PSB region. We define the base of the bias triangles as the zero detuning condition ($\epsilon = 0$) as indicated by the solid green line in Fig. 1(c). Here, the inter-dot level detuning $\epsilon$ refers to the difference in energy levels between the two quantum dots in the system. We measure the microwave response by varying the applied microwave frequency and magnetic field at two different measurement points within the PSB region: low and high detuning points, which are close to and far from the zero-detuning line as indicated by the red circle and green triangle symbols in Fig. 1(c), respectively.

*Contact authors:
sayyid.i.5937@m.isct.ac.jp
kodera.t.a173@m.isct.ac.jp

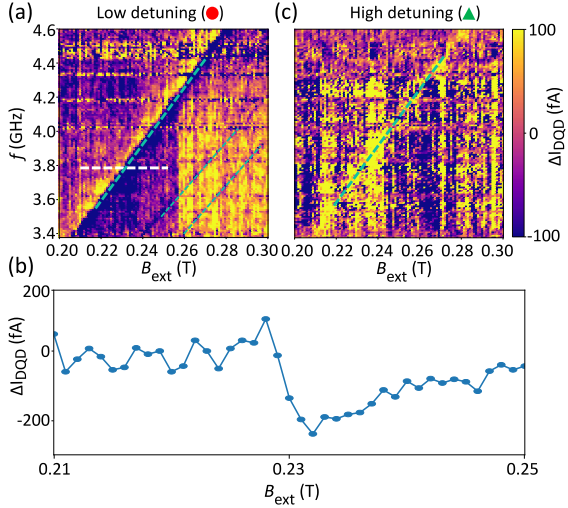

FIG 2. PSB leakage current spectrum measured by sweeping the microwave frequency $f$ and external magnetic field $B_{ext}$. The frequency-dependent average current is subtracted to enhance signal visibility. (a) PSB leakage spectrum at low detuning point, showing three resonance lines, with the main line exhibiting peak-and-dip features. (b) Cross section of the main resonance line at low detuning point at microwave frequency $f$=3.8 GHz as indicated by the white dashed line in (a) showing both peak-and-dip characteristics. (c) PSB leakage at high detuning point, showing a single resonance line with peak characteristic

Figure 2 shows the PSB leakage current spectra obtained at the two measurement points. At the low detuning point (Fig. 2(a)), we observe intriguing results where instead of two peak resonance lines typically expected from PSB leakage driven by EDSR mechanism, we observe three resonance lines. The main resonance line shows an asymmetric peak-and-dip shape, as shown in Fig. 2(b). The other two lines show dip characteristics where the current is lower than the background current. From this result we obtain g factor for the main line is 1.18 and the two dip lines both have g factor of 0.895.

At the high detuning point (Fig. 2(c)), the previously observed two sharp dip lines disappear, and the main resonance line evolves from an asymmetric peak-and-dip shape into a single peak resonance line. The peak resonance line observed here can be readily understood as the result of EDSR mechanism, where a resonant microwave rotates the spins in the DQD, enhancing the PSB leakage current. In a DQD system we expect to observe two resonance lines which correspond to the rotation of left dot spin and right dot spin. However, only a single line is detected in our measurements. This can be attributed to one of the dots having a weaker top gate lever arm and SOC strength, such that the microwave drive cannot rotate the spin in that dot. From the changes of the size and position of the bias triangles on different TG voltages $V_{TG}$, we extract the lever arms of $\alpha_L = 0.05$ eV/V and $\alpha_R = 0.1$ eV/V for the left and right dot. Since $\alpha_R > \alpha_L$ we attribute the observed peak resonance line to the rotation of the right dot spin.

While the EDSR mechanism explains the observation of peak line at high detuning point, it cannot explain the appearance of multiple resonance lines and the asymmetric peak-and-dip resonance at the low detuning point. This suggests the presence of additional mechanisms influencing the system. Another well-known mechanism that can produce resonance lines is MLLZ interference. This mechanism has been shown to generate multiple resonance lines when an oscillating electric field is applied to quantum dot devices with strong SOC and it exhibits a strong dependence on the detuning energy [25,33]. It arises when the AC electric field oscillates the detuning energy through the anticrossings between energy levels, which are formed due to SOC. This process leads to interference between the energy levels, enabling spin transitions between them. Depending on the detuning, the spin resonance lines can appear as either peaks or dips. Although MLLZ interference has been shown to generate multiple resonance lines with either peak or dip shape, it cannot fully describe the asymmetric peak-and-dip shape at low detuning point and the single peak line at high detuning point that we observed in our experiments. This suggests that MLLZ interference alone is insufficient to explain all the features observed in our experimental results.

### III. NUMERICAL SIMULATIONS

To gain further insight, we perform numerical simulations that account for both the EDSR and MLLZ interference effects. Following Ref. [18], we define the Hamiltonian of the two-hole system in a DQD, expressed in the basis of $\{|\uparrow\uparrow\rangle, |\uparrow\downarrow\rangle, |\downarrow\uparrow\rangle, |\downarrow\downarrow\rangle, |S(2,0)\rangle\}$ as

*Contact authors:
sayyid.i.5937@m.isct.ac.jp
kodera.t.a173@m.isct.ac.jp

$$H_{\text{DQD}} = \begin{pmatrix} \frac{\Sigma E_z}{2} & 0 & 0 & 0 & t_\alpha^* \\ 0 & \frac{\delta E_z}{2} & 0 & 0 & t_\beta \\ 0 & 0 & -\frac{\delta E_z}{2} & 0 & -t_\beta^* \\ 0 & 0 & 0 & -\frac{\Sigma E_z}{2} & t_\alpha \\ t_\alpha & t_\beta^* & -t_\beta & t_\alpha^* & -\epsilon \end{pmatrix} \quad (1)$$

where the Zeeman splitting terms $\Sigma E_z$ and $\delta E_z$ correspond to the average and the difference of the Zeeman energies in the two dots, which are formulated as

$$\Sigma E_z = (g_L + g_R)\mu_B B_{\text{ext}} \quad (2)$$
$$\delta E_z = (g_L - g_R)\mu_B B_{\text{ext}} \quad (3)$$

with $g_R$ and $g_L$ denoting the g-factors in the right and left dots respectively, $\mu_B$ the Bohr magneton and $B_{\text{ext}}$ the applied DC magnetic field. The parameters $t_\alpha$ and $t_\beta$ are complex tunnel couplings defined as

$$t_\alpha = -t_y + it_x \quad (4)$$
$$t_\beta = -t_c + it_z \quad (5)$$

where $t_c$ corresponds to the spin conserving tunneling and $t_x, t_y, t_z$ correspond to the component of SOC spin flip tunneling on the convenient basis of orthonormal unpolarized triplet states. In the absence of SOC, the parameter $t_\alpha$ is zero and $t_\beta$ is a real value.

Figure 3 shows the illustration of the energy level diagram of two holes in a DQD as a function of detuning $\epsilon$. The SOC results in the anticrossings between polarized triplet states $|\uparrow\uparrow\rangle, |\downarrow\downarrow\rangle$ and the singlet state $|S(2,0)\rangle$ near the zero detuning. We include the time-dependent Hamiltonian that accounts for the effective oscillating magnetic field driving responsible for the EDSR mechanism $H_{\text{EDSR}}$ and the energy detuning oscillation responsible for the MLLZ interference mechanism $H_{\text{MLLZ}}$, as:

$$H_{\text{EDSR}} = g_R \mu_B B_{\text{eff}} \cos(\omega t)(|\uparrow\uparrow\rangle\langle\uparrow\downarrow| + |\downarrow\downarrow\rangle\langle\downarrow\uparrow| + \text{H. c.}) \quad (6)$$
$$H_{\text{MLLZ}} = \alpha E_{\text{eff}} \cos(\omega t)(|S(2,0)\rangle\langle S(2,0)|) \quad (7)$$

where $g_R\mu_B B_{\text{eff}}$ is the effective amplitude of the EDSR drive, $\alpha E_{\text{eff}}$ is the effective amplitude of the energy detuning oscillation, and $\omega$ is the angular frequency of the applied microwave. Note that we observe only a single peak line at a high detuning point which corresponds to the rotation of the spin on the right dot, therefore we assume that the effective oscillating magnetic field is strongly coupled to the spin in the right dot and neglect the EDSR drive Hamiltonian term on the left dot.

In order to simulate the DQD current, we incorporate the incoherent process terms adapted from Ref. [33], see also Fig. 3(b). These terms are characterized by four distinct tunnel rates $\Gamma$, $\Gamma_{\text{charge}}$, $\Gamma_{\text{spin}}$ and $\Gamma_S$. $\Gamma$ gives the overall tunnelling rate from the (1,0) charge, where there is a hole with a random spin in the left dot to one of the four possible spin states in the (1,1) charge configuration. The relaxation between the spin states in the (1,1) charge configuration is defined as $\Gamma_{\text{spin}}$. $\Gamma_{\text{charge}}$ defines the tunneling process from the unblocked states $|\downarrow\uparrow\rangle$ and $|\uparrow\downarrow\rangle$ to the left dot occupying the $|S(2,0)\rangle$ while the tunneling from $|\downarrow\downarrow\rangle$ and $|\uparrow\uparrow\rangle$ states to $|S(2,0)\rangle$ is blocked due to the PSB effect. $\Gamma_S$ (not shown in the figure) defines charge decoherence process of the $|S(2,0)\rangle$. We then let the system density operator evolve under the Lindblad master equation formulated as

$$\begin{aligned}\frac{d\rho}{dt} &= -i[H_{\text{tot}},\rho] + \Gamma\mathcal{D}[|(1,0)\rangle\langle S(2,0)|]\rho \\ &+ \frac{\Gamma}{4}\mathcal{D}[|S(1,1)\rangle\langle(1,0)|]\rho \\ &+ \sum_j \frac{\Gamma}{4}\mathcal{D}[|T_j(1,1)\rangle\langle(1,0)|]\rho \\ &+ \sum_j \Gamma_{\text{spin}}\mathcal{D}[|S(1,1)\rangle\langle T_j(1,1)|]\rho \\ &+ \Gamma_S\mathcal{D}[|S(2,0)\rangle\langle S(2,0)|]\rho \\ &+ \Gamma_{\text{charge}}\mathcal{D}[|S(1,1)\rangle\langle S(2,0)|]\rho \quad (9)\end{aligned}$$

with

$$H_{\text{tot}} = H_{\text{DQD}} + H_{\text{EDSR}} + H_{\text{MLLZ}} \quad (10)$$

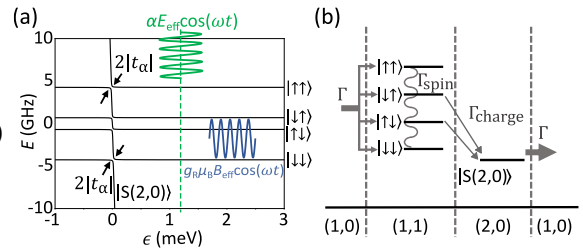

FIG 3. (a) Energy diagram of two holes inside a double quantum dot. Anticrossings between the energy levels originate from spin flip tunnel coupling due to SOC. The effective oscillating magnetic field rotates the spin in the right dot as illustrated by the blue wave. The electric AC signal oscillates the detuning position as illustrated by the green wave. (b) Schematic of the

*Contact authors:
sayyid.i.5937@m.isct.ac.jp
kodera.t.a173@m.isct.ac.jp

transport cycle used in the simulation which includes the relevant incoherent processes. We assume that the tunnel rates into and out of the dot are equal and given by Γ. Immediately after one of the holes tunnel out of the dot, a hole randomly tunnels into the dot, occupying one of the four possible states in (1,1) charge configuration with equal probability.

Where $\mathcal{D}[A]\rho = -1/2\{A^\dagger A, \rho\} + A\rho A^\dagger$ is the Lindblad super operator representing the incoherent processes and the $T_j$ corresponds to the three triplet state $\{T_+, T_-, T_0\}$. We initialize the system in the $|\downarrow\downarrow\rangle$ state and let the system evolve for several cycles until it reaches the steady state condition. The DQD current is calculated as $I_{DQD} = e\Gamma p_{10}$ where $p_{10}$ is the steady state population of the state (1,0). For the simulation, we fix the parameters $g_R$=1.18 and $g_L$=0.895 obtained from the experimental results. We can reproduce the experimental results in the simulation with the following parameters: $t_x = 0.5$ GHz, $t_y = 0.5$ GHz, $t_z = 0.1$ GHz, $t_c = 0.8$ GHz, $\Gamma = 200$ GHz, $\Gamma_{charge} = 1$ GHz, $\Gamma_S = 10$ GHz, $\Gamma_{spin} = 1$ MHz, $g_R\mu_B B_{eff} = 8.2$ neV, and $\alpha E_{eff} = 0.33$ meV. For clarity, the current in the simulation results is adjusted by subtracting the background current obtained when the system is driven in a non-resonance condition.

Figure 4(a) shows the simulation result at the low detuning point where $\epsilon = 0.2$ meV. Three resonance dip lines are obtained with the main line accompanied by a peak feature, consistent with the experimental results. We interpret the peak-and-dip feature of the main line as arising from the detuning dependent efficiencies of the transitions labelled as $E_{1E}$ and $E_{1M}$, which are induced by the EDSR and MLLZ mechanisms, respectively. The latter mechanism is enhanced in the low detuning regime when the detuning oscillates across the zero-detuning anticrossings due to a large microwave signal. When the driving frequency is on resonance with a pair of energy levels in the (1,1) charge configuration, the MLLZ interference enables spin transition among the two energy levels and the $|S(2,0)\rangle$ state. The resonance shape resulted due to this interference effect can be a peak or a dip depending on the net transition rate [32]. In the configuration of $\epsilon = 0.2$ meV, the transition from the $|S(2,0)\rangle$ state to the $|\downarrow\downarrow\rangle$ and $|\downarrow\uparrow\rangle$ is dominant leading to a dip feature. The peak feature can be explained as the result of EDSR which rotates the spin in the right dot leading to transition between the $|\downarrow\downarrow\rangle$ and $|\downarrow\uparrow\rangle$ states. This transition is indicated by the arrows labeled as $E_{1E}$ in Figs. 4(a) and 4(c).

We interpret the other two dip lines obtained in the simulation as the transition of the left dot spin $|\downarrow\downarrow\rangle \leftrightarrow |\uparrow\downarrow\rangle$, and two photon process transition $|\uparrow\uparrow\rangle \leftrightarrow |\downarrow\downarrow\rangle$, indicated by $E_2$ and $E_{T/2}$ respectively, in Fig. 4(b) and Fig. 4(c). The $E_2$ transition originates from the MLLZ interference similar to the $E_{1M}$ transition process while the two photon process $E_{T/2}$ corresponds to a resonance condition of dark Bell states generation, characterized by a sharp dip appearing at the average Zeeman splitting frequency [41]. This phenomenon is caused by the interplay of the spin transition between the blocked states and coherent interdot tunneling whose effect is enhanced near zero detuning conditions. In our case, the spin transition between the two blocked states is mediated by the MLLZ interference.

Figure 4(b) shows the simulation result at the high detuning point where $\epsilon = 2$ meV. A single peak resonance line is obtained consistent to the experimental results. In this condition, $\alpha E_{eff} < \epsilon$, which suppresses the MLLZ interference effect. The observed peak resonance line can therefore be attributed to the EDSR mechanism, as explained in the low detuning case. This transition is indicated by the arrows labeled as $E_{1E}$ in Figs. 4(b) and 4(c). With these interpretations we explain all the features observed in the experimental results.

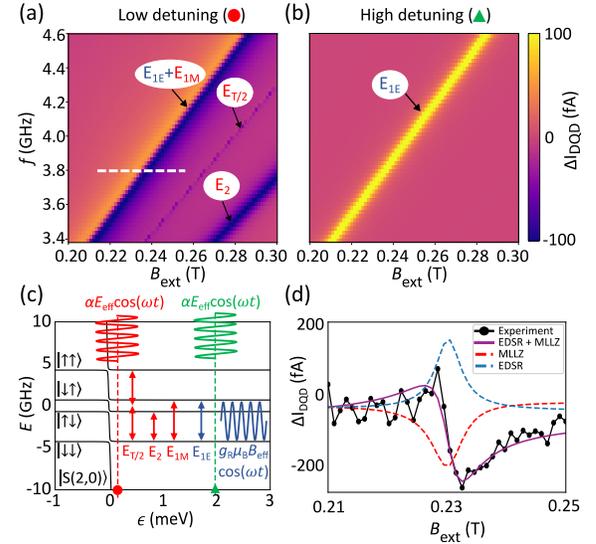

FIG 4. Simulation results of PSB leakage current spectrum at (a) the low detuning condition ($\epsilon = 0.2$ meV). (b) the high detuning condition ($\epsilon = 2$ meV). Each resonance line is associated with a transition between the energy levels indicated in (c). (c) Schematic illustration of microwave driving at high detuning (green wave) and low detuning (red wave). The red arrows denote transitions induced by MLLZ interference at low detuning, while the blue arrow denotes the transition due to EDSR drive (blue wave).

*Contact authors:
sayyid.i.5937@m.isct.ac.jp
kodera.t.a173@m.isct.ac.jp

(d) Comparison between the simulation and the experimental data from Fig. 2(c) at a microwave frequency $f = 3.8$ GHz. Good agreement is obtained only when both EDSR and MLLZ effects are included in the simulation.

To evaluate the contribution of each mechanism to the resonance line shape, we perform a least square fit of the numerical model to the peak-and-dip resonance line obtained at low detuning point at a fixed frequency of 3.8 GHz, as shown in Fig. 4(d). We fit the model by considering the presence of both EDSR and MLLZ interference, as well as the case where one of them is neglected by setting the corresponding Hamiltonian parameters to zero. We find that the asymmetric peak-and-dip shape can be accurately fitted only when both EDSR and MLLZ interference are included. In contrast, disabling either mechanism results in a poor fit to the experimental data, yielding only a peak or dip shape, corresponding to the EDSR mechanism and MLLZ interference, respectively. These results establish that both EDSR and MLLZ are the key mechanisms underlying the multiple resonance lines with asymmetric peak–dip shapes observed in the experiments.

## IV. DISCUSSION

We have shown evidence of the coexistence of EDSR and MLLZ through the qualitative agreement between the experimental and simulation results. However, there are some features that fail to be reproduced. In what follows, we discuss possible explanations for the discrepancies and explore other potential mechanisms that might produce similar results. We then present our reasoning that the interplay between EDSR and MLLZ is the primary mechanism at work.

We begin by discussing the possible origins of the discrepancies between simulation and experimental results. In particular, the widths and slopes of the two sharp dip resonance lines are not quantitatively reproduced by our simulations. In the experimental data, one of the sharp dip lines does not necessarily originate from a transition of the left-dot hole spin mediated by the MLLZ mechanism (the $E_2$ transition). Instead, it may arise from the formation of dark Bell states involving other coupled spin pairs in the system. Possible candidates include a hole spin in the right dot coupled to a spin in an excited state of the left dot, or coupling to hole spins in nearby parasitic dots. Such scenarios could account for the observed sharp resonance lines, as dark Bell states are expected to exhibit reduced decoherence compared with single-spin states. Within this interpretation, the two sharp dip resonance lines may correspond to two distinct dark Bell states: one formed by coupling between the ground-state hole spins in the right and left dots, and the other formed by coupling between the ground-state hole spin in the right dot and an excited-state hole spin in the left dot. In this framework, an effective g factor of the left-dot spin on the order of 0.5 would be required to reproduce the slopes observed in Fig 2(c).

We note that an asymmetric spectral line shape can arise from MLLZ interference alone (i.e., without EDSR), within a certain detuning range, see Fig. 6 of Ref. [27]. However, the polarity differs; in our case, the peak structure is observed at a lower magnetic field, whereas in the pure MLLZ case, it would be the opposite. Moreover, the MLLZ interference alone cannot explain the observation of a peak line at high detuning point. Another possible mechanism for the asymmetric spectral line shape is an effective oscillating magnetic field along the spin quantization axis (i.e. modulation of the Zeeman splitting) [27]. However, this cannot account for the generation of multiple resonance lines and simulations incorporating this effect fail to fit the observed peak-and-dip shape. We therefore conclude that the asymmetric spectral line shape primarily originates from the interplay between EDSR and MLLZ interference.

In summary, this work reports unconventional spin resonance features that depend on the energy detuning. At the high detuning point, we observe a single resonance peak line, while at the low detuning point, we observe multiple dip resonance lines, one of which exhibits asymmetric peak-and-dip shape. These features are well explained by the interplay between EDSR and MLLZ interference when electrical microwaves are applied in the presence of SOC, as verified by numerical simulations. Although some discrepancies between the experimental and simulation results remain, our results highlight the importance of this interplay, which may complicate the dynamics of electrically driven spin qubits. An important direction for future work is to investigate how this interplay affects spin-qubit control. Such effects may become particularly relevant when applying microwaves near zero detuning, such as in the implementation of two-qubit controlled-rotation gates [7,22].

## ACKNOWLEDGMENTS
This research was financially supported by JST Moonshot R&D grant number JPMJMS2065, MEXT Quantum Leap Flagship Program (MEXT QLEAP) grant number JPMXS0118069228, JST FOREST grant number JPMJFR244D, JST ASPIRE grant number JPMJAP25A2, and Grants-in-Aid for

*Contact authors:
sayyid.i.5937@m.isct.ac.jp
kodera.t.a173@m.isct.ac.jp

Scientific Research grant numbers JP23H05455, JP23K26483, and JP23K17327.

**DATA AVAILABILITY**

The data that support the findings of this article are not publicly available. The data are available from the authors upon reasonable request.

[1] M. F. Gonzalez-Zalba, S. de Franceschi, E. Charbon, T. Meunier, M. Vinet, and A. S. Dzurak, Scaling silicon-based quantum computing using CMOS technology, Nat. Electron. **4**, 872 (2021).

[2] A. Laucht, F. Hohls, N. Ubbelohde, M. Fernando Gonzalez-Zalba, D. J. Reilly, S. Stobbe, T. Schröder, P. Scarlino, J. V. Koski, A. Dzurak, C.-H. Yang, J. Yoneda, F. Kuemmeth, H. Bluhm, J. Pla, C. Hill, J. Salfi, A. Oiwa, J. T. Muhonen, E. Verhagen, M. D. LaHaye, H. H. Kim, A. W. Tsen, D. Culcer, A. Geresdi, J. A. Mol, V. Mohan, P. K. Jain, and J. Baugh, Roadmap on quantum nanotechnologies, Nanotechnology **32**, 162003 (2021).

[3] D. Rotta, F. Sebastiano, E. Charbon, and E. Prati, Quantum information density scaling and qubit operation time constraints of CMOS silicon-based quantum computer architectures, Npj Quantum Inf. **3**, 26 (2017).

[4] N. P. de Leon, K. M. Itoh, D. Kim, K. K. Mehta, T. E. Northup, H. Paik, B. S. Palmer, N. Samarth, S. Sangtawesin, and D. W. Steuerman, Materials challenges and opportunities for quantum computing hardware, Science **372**, eabb2823 (2021).

[5] N. W. Hendrickx, W. I. L. Lawrie, M. Russ, F. van Riggelen, S. L. de Snoo, R. N. Schouten, A. Sammak, G. Scappucci, and M. Veldhorst, A four-qubit germanium quantum processor, Nature **591**, 580 (2021).

[6] L. C. Camenzind, S. Geyer, A. Fuhrer, R. J. Warburton, D. M. Zumbühl, and A. V. Kuhlmann, A hole spin qubit in a fin field-effect transistor above 4 kelvin, Nat. Electron. **5**, 178 (2022).

[7] N. W. Hendrickx, D. P. Franke, A. Sammak, G. Scappucci, and M. Veldhorst, Fast two-qubit logic with holes in germanium, Nature **577**, 487 (2020).

[8] R. Maurand, X. Jehl, D. Kotekar-Patil, A. Corna, H. Bohuslavskyi, R. Laviéville, L. Hutin, S. Barraud, M. Vinet, M. Sanquer, and S. De Franceschi, A CMOS silicon spin qubit, Nat. Commun. **7**, 13575 (2016).

[9] W. Huang, C. H. Yang, K. W. Chan, T. Tanttu, B. Hensen, R. C. C. Leon, M. A. Fogarty, J. C. C. Hwang, F. E. Hudson, K. M. Itoh, A. Morello, A. Laucht, and A. S. Dzurak, Fidelity benchmarks for two-qubit gates in silicon, Nature **569**, 532 (2019).

[10] R. Zhao, T. Tanttu, K. Y. Tan, B. Hensen, K. W. Chan, J. C. C. Hwang, R. C. C. Leon, C. H. Yang, W. Gilbert, F. E. Hudson, K. M. Itoh, A. A. Kiselev, T. D. Ladd, A. Morello, A. Laucht, and A. S. Dzurak, Single-spin qubits in isotopically enriched silicon at low magnetic field, Nat. Commun. **10**, 5500 (2019).

[11] J. Yoneda, K. Takeda, T. Otsuka, T. Nakajima, M. R. Delbecq, G. Allison, T. Honda, T. Kodera, S. Oda, Y. Hoshi, N. Usami, K. M. Itoh, and S. Tarucha, A quantum-dot spin qubit with coherence limited by charge noise and fidelity higher than 99.9%, Nat. Nanotechnol. **13**, 102 (2018).

[12] M. Veldhorst, C. H. Yang, J. C. C. Hwang, W. Huang, J. P. Dehollain, J. T. Muhonen, S. Simmons, A. Laucht, F. E. Hudson, K. M. Itoh, A. Morello, and A. S. Dzurak, A two-qubit logic gate in silicon, Nature **526**, 410 (2015).

[13] F. H. L. Koppens, C. Buizert, K. J. Tielrooij, I. T. Vink, K. C. Nowack, T. Meunier, L. P. Kouwenhoven, and L. M. K. Vandersypen, Driven coherent oscillations of a single electron spin in a quantum dot, Nature **442**, 766 (2006).

[14] M. Veldhorst, H. G. J. Eenink, C. H. Yang, and A. S. Dzurak, Silicon CMOS architecture for a spin-based quantum computer, Nat. Commun. **8**, 1766 (2017).

[15] M. A. Fogarty, K. W. Chan, B. Hensen, W. Huang, T. Tanttu, C. H. Yang, A. Laucht, M. Veldhorst, F. E. Hudson, K. M. Itoh, D. Culcer, T. D. Ladd, A. Morello, and A. S. Dzurak, Integrated silicon qubit platform with single-spin addressability, exchange control and single-shot singlet-triplet readout, Nat. Commun. **9**, 17 (2018).

[16] C. H. Yang, R. C. C. Leon, J. C. C. Hwang, A. Saraiva, T. Tanttu, W. Huang, J. Camirand Lemyre, K. W. Chan, K. Y. Tan, F. E. Hudson, K. M. Itoh, A. Morello, M. Pioro-Ladrière, A. Laucht, and A. S. Dzurak, Operation of a silicon quantum processor unit cell above one kelvin, Nature **580**, 350 (2020).

[17] J. Y. Huang, R. Y. Su, W. H. Lim, M. Feng, B. Van Straaten, B. Severin, W. Gilbert, N. Dumoulin Stuyck, T. Tanttu, S. Serrano, J. D. Cifuentes, I. Hansen, A. E. Seedhouse, E.

*Contact authors:
sayyid.i.5937@m.isct.ac.jp
kodera.t.a173@m.isct.ac.jp

Vahapoglu, R. C. C. Leon, N. V. Abrosimov, H.-J. Pohl, M. L. W. Thewalt, F. E. Hudson, C. C. Escott, N. Ares, S. D. Bartlett, A. Morello, A. Saraiva, A. Laucht, A. S. Dzurak, and C. H. Yang, High-fidelity spin qubit operation and algorithmic initialization above 1 K, Nature **627**, 772 (2024).

[18] J. Danon and Yu. V. Nazarov, Pauli spin blockade in the presence of strong spin-orbit coupling, Phys. Rev. B **80**, 041301 (2009).

[19] J. R. Petta, A. C. Johnson, J. M. Taylor, E. A. Laird, A. Yacoby, M. D. Lukin, C. M. Marcus, M. P. Hanson, and A. C. Gossard, Coherent Manipulation of Coupled Electron Spins in Semiconductor Quantum Dots, Science **309**, 2180 (2005).

[20] G. Yamahata, T. Kodera, H. O. H. Churchill, K. Uchida, C. M. Marcus, and S. Oda, Magnetic field dependence of Pauli spin blockade: A window into the sources of spin relaxation in silicon quantum dots, Phys. Rev. B **86**, 115322 (2012).

[21] A. E. Seedhouse, T. Tanttu, R. C. C. Leon, R. Zhao, K. Y. Tan, B. Hensen, F. E. Hudson, K. M. Itoh, J. Yoneda, C. H. Yang, A. Morello, A. Laucht, S. N. Coppersmith, A. Saraiva, and A. S. Dzurak, Pauli Blockade in Silicon Quantum Dots with Spin-Orbit Control, PRX Quantum **2**, 010303 (2021).

[22] S. Geyer, B. Hetényi, S. Bosco, L. C. Camenzind, R. S. Eggli, A. Fuhrer, D. Loss, R. J. Warburton, D. M. Zumbühl, and A. V. Kuhlmann, Anisotropic exchange interaction of two hole-spin qubits, Nat. Phys. **20**, 1152 (2024).

[23] S. Nadj-Perge, V. S. Pribiag, J. W. G. Van Den Berg, K. Zuo, S. R. Plissard, E. P. A. M. Bakkers, S. M. Frolov, and L. P. Kouwenhoven, Spectroscopy of Spin-Orbit Quantum Bits in Indium Antimonide Nanowires, Phys. Rev. Lett. **108**, 166801 (2012).

[24] E. A. Laird, C. Barthel, E. I. Rashba, C. M. Marcus, M. P. Hanson, and A. C. Gossard, A new mechanism of electric dipole spin resonance: hyperfine coupling in quantum dots, Semicond. Sci. Technol. **24**, 064004 (2009).

[25] J. Stehlik, M. D. Schroer, M. Z. Maialle, M. H. Degani, and J. R. Petta, Extreme harmonic generation in electrically driven spin resonance, Phys. Rev. Lett. **112**, 227601 (2014).

[26] A. Crippa, R. Maurand, L. Bourdet, D. Kotekar-Patil, A. Amisse, X. Jehl, M. Sanquer, R. Laviéville, H. Bohuslavskyi, L. Hutin, S. Barraud, M. Vinet, Y. M. Niquet, and S. De Franceschi, Electrical Spin Driving by g-Matrix Modulation in Spin-Orbit Qubits, Phys. Rev. Lett. **120**, 137702 (2018).

[27] A. Sala and J. Danon, Line shapes of electric dipole spin resonance in Pauli spin blockade, Phys. Rev. B **104**, 085421 (2021).

[28] N. Ares, G. Katsaros, V. N. Golovach, J. J. Zhang, A. Prager, L. I. Glazman, O. G. Schmidt, and S. De Franceschi, SiGe quantum dots for fast hole spin Rabi oscillations, Appl. Phys. Lett. **103**, 263113 (2013).

[29] K. C. Nowack, F. H. L. Koppens, Y. V. Nazarov, and L. M. K. Vandersypen, Coherent control of a single electron spin with electric fields, Science **318**, 1430 (2007).

[30] S. Nadj-Perge, S. M. Frolov, E. P. A. M. Bakkers, and L. P. Kouwenhoven, Spin–orbit qubit in a semiconductor nanowire, Nature **468**, 1084 (2010).

[31] J. W. G. van den Berg, S. Nadj-Perge, V. S. Pribiag, S. R. Plissard, E. P. A. M. Bakkers, S. M. Frolov, and L. P. Kouwenhoven, Fast Spin-Orbit Qubit in an Indium Antimonide Nanowire, Phys. Rev. Lett. **110**, 066806 (2013).

[32] J. Danon and M. S. Rudner, Multilevel interference resonances in strongly driven three-level systems, Phys. Rev. Lett. **113**, 247002 (2014).

[33] J. Stehlik, M. Z. Maialle, M. H. Degani, and J. R. Petta, Role of multilevel Landau-Zener interference in extreme harmonic generation, Phys. Rev. B **94**, 075307 (2016).

[34] K. Horibe, T. Kodera, and S. Oda, Lithographically defined few-electron silicon quantum dots based on a silicon-on-insulator substrate, Appl. Phys. Lett. **106**, 083111 (2015).

[35] Y. Yamaoka, S. Oda, and T. Kodera, Electron transport in physically-defined double quantum dots on a highly doped silicon-on-insulator substrate, Appl. Phys. Lett. **109**, 113109 (2016).

[36] S. Mizoguchi, N. Shimatani, M. Kobayashi, T. Makino, Y. Yamaoka, and T. Kodera, Fabrication and characterization of physically defined quantum dots on a boron-doped silicon-on-insulator substrate, Jpn. J. Appl. Phys. **57**, 04FK03 (2018).

[37] S. Nishiyama, K. Kato, Y. Liu, R. Mizokuchi, J. Yoneda, T. Kodera, and T. Mori, Single-Electron Transistor Operation of a Physically Defined Silicon Quantum Dot Device Fabricated by Electron Beam Lithography Employing a Negative-Tone Resist, IEICE Trans. Electron. **E106.C**, 592 (2023).

*Contact authors:
sayyid.i.5937@m.isct.ac.jp
kodera.t.a173@m.isct.ac.jp

[38] G. Yamahata, T. Kodera, H. O. H. Churchill, K. Uchida, C. M. Marcus, and S. Oda, Magnetic field dependence of Pauli spin blockade: A window into the sources of spin relaxation in silicon quantum dots, Phys. Rev. B **86**, 115322 (2012).

[39] H. Wei, S. Mizoguchi, R. Mizokuchi, and T. Kodera, Estimation of hole spin g-factors in p-channel silicon single and double quantum dots towards spin manipulation, Jpn. J. Appl. Phys. **59**, SGGI10 (2020).

[40] R. Mizokuchi, S. Oda, and T. Kodera, Physically defined triple quantum dot systems in silicon on insulator, Appl. Phys. Lett. **114**, 073104 (2019).

[41] R. Sánchez and G. Platero, Dark Bell states in tunnel-coupled spin qubits, Phys. Rev. B **87**, 081305 (2013).

*Contact authors:
sayyid.i.5937@m.isct.ac.jp
kodera.t.a173@m.isct.ac.jp